\documentclass[final,twoside,11pt]{entics} 
\usepackage{enticsmacro}
\usepackage{graphicx}
\usepackage[all]{xy}

\usepackage{todonotes}
\usepackage{wrapfig}
\usepackage{mathrsfs}  
\usepackage{stmaryrd}
\usepackage{amsmath}
\usepackage{mathtools}
\usepackage{amsfonts}
\usepackage{amssymb}
\usepackage{amsopn}
\usepackage{cmll}
\usepackage{framed}
\usepackage{wasysym}
\usepackage{hyperref}
\usepackage{cleveref}
\crefname{subsubsection}{Sec.}{Sec.}
\crefname{subsection}{Sec.}{Sec.}
\crefname{figure}{Fig.}{Fig.}
\crefname{section}{Sec.}{Sec.}

\usepackage{tikz}
\usetikzlibrary{arrows}
\usetikzlibrary{decorations.pathmorphing}
\usetikzlibrary{decorations.markings}
\usetikzlibrary{calc}
\usetikzlibrary{backgrounds}
\tikzstyle{data}=[->,>=stealth, densely dashed, thick, darkbrown]
\tikzstyle{control}=[->, thick,  -open triangle 60]

\tikzset{
  prof/.style = {decoration = {markings, mark = at position 0.5 with { \node[transform shape, yscale=.4] {$|$}; } }, postaction = {decorate} },
}
\tikzstyle{conflict}=[snake it,-]
\tikzstyle{coincidence}=[thick]
\tikzstyle{internal}=[draw]
\tikzstyle{positive}[2]=[draw, circle, thick, fill=blue, label={[xshift=-0.3em, yshift=0.2em]below right:\scalebox{0.6}{$#1$}}]
\tikzstyle{negative}[2]=[draw, circle, thick, fill=red, label={[xshift=-0.3em, yshift=0.2em]below right:\scalebox{0.6}{$#1$}}]
\tikzstyle{neutral}[2]=[draw, circle, thick, fill=gray, label={[xshift=-0.3em, yshift=0.2em]below right:\scalebox{0.6}{$#1$}}]

\tikzstyle{background rectangle}=
  [fill=gray!20]

\tikzset{snake it/.style={decorate, decoration={snake, amplitude=.3mm,segment length=1mm}}}

\usepackage{bm}
\usepackage{tikz-cd}

\usepackage{bussproofs} %

\definecolor{codegreen}{rgb}{0,0.6,0}
\definecolor{codegray}{rgb}{0.5,0.5,0.5}
\definecolor{codepurple}{rgb}{0.58,0,0.82}

\usepackage{scalerel}
\newsavebox{\lXbrace}
\savebox{\lXbrace}{$\llparenthesis$}
\newsavebox{\rXbrace}
\savebox{\rXbrace}{$\rrparenthesis$}

\newcommand{\Conf}[1]{\mathscr{C}\hspace{-1pt}(#1)}
\renewcommand{\pol}{\text{pol}}

\newcommand{\sym}[1]{\mathcal{S}(#1)}
\newcommand{\symgp}[2]{\mathcal{S}_{#1}(#2)}

\newcommand{\negsym}[1]{\mathcal{S}_-(#1)}
\newcommand{\possym}[1]{\mathcal{S}_+(#1)}

\newcommand{\lift}[1]{\hat{#1}}

\newcommand{\cc}{{\mathrm {c\!c}}}

\newcommand{\Set}{\mathbf{Set}}

\def\profto{\mathrel{\mkern3mu  \vcenter{\hbox{$\scriptscriptstyle+$}}%
                    \mkern-12mu{\to}}}

\newcommand\restr[2]{{\left.\kern-\nulldelimiterspace #1 \vphantom{\big|} \right|_{#2} }}

\usepackage{mathtools}
\usepackage{enumitem}
\usepackage{adjustbox}

\newcommand{\id}{\mathrm{id}}

 \definecolor{darkbrown}{rgb}{0.4, 0.26, 0.2}
\newcommand{\imc}{\rightarrowtriangle}

\newcommand{\conflict}{\mathrel{\#}}

\newcommand{\act}{\cdot}

\newcommand{\negauto}[1]{\mathcal{N}_{#1}}
\newcommand{\posauto}[1]{\mathcal{P}_{#1}}

\newcommand{\Aut}[1]{\mathrm{Aut}(#1)}
\newcommand{\display}[1]{p_{#1}}

\renewcommand{\neg}{\mathcal{N}}
\newcommand{\pos}{\mathcal{P}}

\newcommand{\nact}{\mathit{act}_\neg}
\newcommand{\pact}{\mathit{act}_\pos}

\tikzset{curve/.style={settings={#1},to path={(\tikztostart)
    .. controls ($(\tikztostart)!\pv{pos}!(\tikztotarget)!\pv{height}!270:(\tikztotarget)$)
    and ($(\tikztostart)!1-\pv{pos}!(\tikztotarget)!\pv{height}!270:(\tikztotarget)$)
    .. (\tikztotarget)\tikztonodes}},
    settings/.code={\tikzset{quiver/.cd,#1}
        \def\pv##1{\pgfkeysvalueof{/tikz/quiver/##1}}},
    quiver/.cd,pos/.initial=0.35,height/.initial=0}

\def\conf{MFPS 2022} 	
\volume{1}			
\def\volu{1}			
\def\papno{13}			
\def\mydoi{{\normalshape \href{https://doi.org/10.46298/entics.10498}{10.46298/entics.10498}}}
\def\copynames{H. Paquet} 
\def\runtitle{Bi-invariance for Uniform Strategies on Event Structure}
\def\lastname{Paquet}
\begin{document}
\begin{frontmatter}
  \title{Bi-invariance for Uniform Strategies on Event Structures} 
  \author{Hugo Paquet\thanksref{myemail}}
  \address{Department of Computer Science\\ University of Oxford\\          
    UK}                            
   \thanks[myemail]{Email: \href{mailto:hugo.paquet@cs.ox.ac.uk} {\texttt{\normalshape
        hugo.paquet@cs.ox.ac.uk}}} 
\begin{abstract} 
A recurring problem in game semantics is to enforce uniformity in strategies. Informally, a strategy is uniform when the Player's behaviour does not depend on the particular indexing of moves chosen by the Opponent. In game semantics, uniformity is used to define a resource modality $\oc$, that can be exploited for the semantics of programming languages.

In this paper we give a new account of uniformity for strategies on event structures.  This work is inspired by an older idea by Melliès, that uniformity should be expressed as ``bi-invariance'' with respect to two interacting group actions. 
We explore the algebraic foundations of bi-invariance, adapt this idea to the language of event structures and define a general notion of uniform strategy in this context. 
Finally we revisit the existing approach to uniformity, and show how this arises as a special case of our constructions. 
\end{abstract}
\end{frontmatter}

\section{Introduction}
\label{sec:intro}

This paper is about the foundations of game semantics, and in
particular the notion of uniformity for strategies. Informally, uniformity is the property that a strategy does not depend on the particular indexing of moves by the Opponent. This is a key component in various kinds of game semantics \cite{abramsky2000full,castellan2014symmetry,mellies2007asynchronous}, necessary to obtain a cartesian closed model and thus validate basic soundness properties.  

In this paper we explore an algebraic description of uniformity, inspired by an unpublished manuscript by Melliès (\cite{mellies2003asynchronous}, 2003). Melliès argues for a treatment of uniformity as ``bi-invariance'' under two interacting group actions. We take this idea further: 
\begin{itemize}
\item we give an abstract definition of bi-invariance in categorical terms;
\item we propose a new, general notion of uniform strategy based on event structures; and 
\item we formally relate this to the state-of-the-art approach to uniformity in concurrent games \cite{cg2}, shedding new light on existing definitions of uniformity.
\end{itemize}

In \S\ref{sec:intro:intro} we give an informal introduction to uniformity, using some examples of uniform and non-uniform strategies for a simple concrete game. Then we introduce event structures, games and strategies (\S\ref{sec:intro:cg}), and we formally motivate uniformity and bi-invariance (\S\ref{sec:intro:uniformity}).

\subsection{Uniform and non-uniform strategies}
\label{sec:intro:intro}

Consider a two-player game where the only possible action for each player is to put tokens down on the table. Each player has access to an infinite supply of tokens; tokens can be put down at any time and in any order; and the tokens are indistinguishable. So the game consists of 
\begin{equation}
\label{eq:trivialgame}
\raisebox{-1em}{
\begin{tikzpicture} 
\node[positive=0] (1) at (0, 0) {};
\node[positive=1] (1) at (1, 0) {};
\node[positive=2] (1) at (2, 0) {};
\node (1) at (3, 0) {\Huge$\dots$};

\node[negative=0] (1) at (6, 0) {};
\node[negative=1] (1) at (7, 0) {};
\node[negative=2] (1) at (8, 0) {};
\node (1) at (9, 0) {\Huge$\dots$};
\end{tikzpicture}
}
\end{equation}
where Blue and Red are the two players and their respective tokens are indexed by $\mathbb N$. We want to reason about strategies for Blue (regardless of winning conditions for the game). Here are three possible strategies: 
\begin{enumerate}
\item ``Put a token down every time Red puts a token down.''
\label{strat1}
\item ``Never put down any tokens.''
\label{strat2}
\item ``Put a token down without waiting for Red, and \emph{two more} every time Red puts down a token.''
\label{strat3}
\end{enumerate}
And here is a possible graphical representation for each of them: 
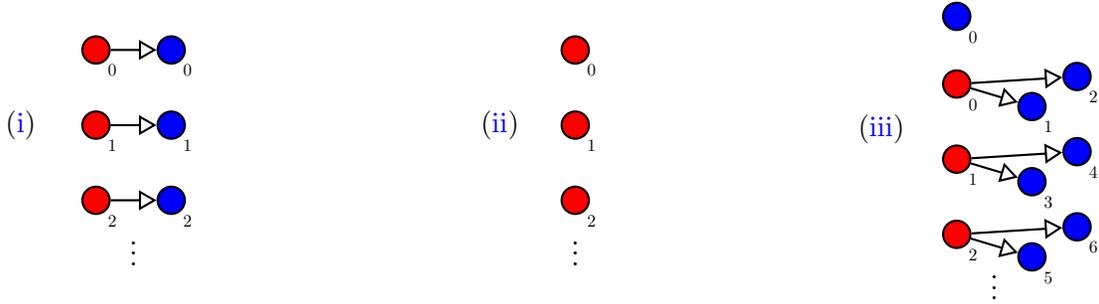
\[
\begin{minipage}{0.33\textwidth}
\centering
\begin{tikzpicture}
\node[negative=0] (0) at (0, 3) {};
\node[negative=1] (1) at (0, 2) {};
\node[negative=2] (2) at (0, 1) {};

\node[positive=0] (0') at (1, 3) {};
\node[positive=1] (1') at (1, 2) {};
\node[positive=2] (2') at (1, 1) {};

\draw[control] (0) to (0');
\draw[control] (1) to (1');
\draw[control] (2) to (2');

\node (item) at (-1, 2) {\eqref{strat1}};
\node (dots) at (0.5, 0.4) {$\vdots$};
\end{tikzpicture}
\end{minipage}%
\begin{minipage}{0.33\textwidth}
\centering
\begin{tikzpicture}
\node[negative=0] (0) at (0, 3) {};
\node[negative=1] (1) at (0, 2) {};
\node[negative=2] (2) at (0, 1) {};

\node (item) at (-1, 2) {\eqref{strat2}};
\node (dots) at (0, 0.4) {$\vdots$};
\end{tikzpicture}
\end{minipage}%
\begin{minipage}{0.33\textwidth}
\centering
\begin{tikzpicture}
\node[negative=0] (0) at (0, 3) {};
\node[negative=1] (1) at (0, 2) {};
\node[negative=2] (2) at (0, 1) {};

\node[positive=1] (0') at (1, 2.7) {};
\node[positive=3] (1') at (1, 1.7) {};
\node[positive=5] (2') at (1, 0.7) {};

\node[positive=2] (01) at (1.6, 3.1) {};
\node[positive=4] (11) at (1.6, 2.1) {};
\node[positive=6] (21) at (1.6, 1.1) {};

\node[positive=0] (alone) at (0, 3.9) {};

\draw[control] (0) to (0');
\draw[control] (1) to (1');
\draw[control] (2) to (2');
\draw[control] (0) to (01);
\draw[control] (1) to (11);
\draw[control] (2) to (21);

\node (item) at (-1, 2.4) {\eqref{strat3}};
\node (dots) at (0.5, 0.4) {$\vdots$};
\end{tikzpicture}
\end{minipage}
\]
Arrows specify the dependency of Blue actions on Red actions. The absence of an arrow between two moves means that they may happen in any order or at the same time. This is true in particular for Red moves: Blue does not know which specific tokens Red will choose to play.

For Blue, the choice of which token to put down at a given point is arbitrary: tokens are indistinguishable. Strategy \eqref{strat1} does not explicitly require that Blue plays the same token number as Red, and Blue's choice does not affect the course of the game. 

But if the choice of token does not matter, then our representation should not allow for the following strategies: 
\begin{enumerate}[resume]
\item ``Put a token down only if Red puts down token number 2.'' \label{strat4}
\item ``Put $n$ tokens down after Red plays token number $n$.''
\label{strat5}
\end{enumerate}
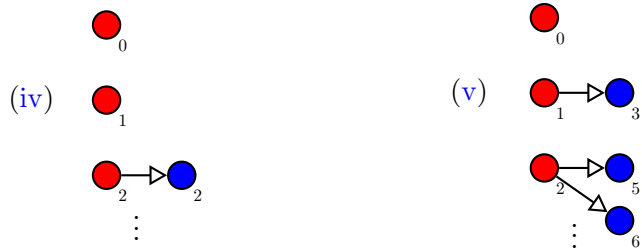
\[
\begin{minipage}{0.33\textwidth}
\centering
\begin{tikzpicture}
\node[negative=0] (0) at (0, 3) {};
\node[negative=1] (1) at (0, 2) {};
\node[negative=2] (2) at (0, 1) {};

\node[positive=2] (2') at (1, 1) {};
\draw[control] (2) to (2');
\node (item) at (-1, 2) {\eqref{strat4}};
\node (dots) at (0.4, 0.4) {$\vdots$};
\end{tikzpicture}
\end{minipage}%
\begin{minipage}{0.33\textwidth}
\centering
\begin{tikzpicture}
\node[negative=0] (0) at (0, 3) {};
\node[negative=1] (1) at (0, 2) {};
\node[negative=2] (2) at (0, 1) {};

\node[positive=3] (1') at (1, 2) {};
\node[positive=5] (2') at (1, 1) {};

\node[positive=6] (21) at (1, 0.3) {};

\draw[control] (2) to (2');
\draw[control] (1) to (1');
\draw[control] (2) to (21);
\node (item) at (-1, 2) {\eqref{strat5}};
\node (dots) at (0.4, 0.2) {$\vdots$};
\end{tikzpicture}
\end{minipage}
\]
We say that these strategies are \textbf{non-uniform}, because the behaviour of Blue is dependent on the token that Red chooses to play. In more general games, uniformity is difficult to express and reason about.  


\subsection{Concurrent games}
\label{sec:intro:cg}

We are interested in a particular mathematical theory of games and strategies, based on event structures, initially developed by Rideau and Winskel \cite{rideau2011concurrent}. These games are often called ``concurrent games'' because of the possibility for players to perform multiple actions at the same time, as in strategy \eqref{strat3} above. 

Concurrent games can be presented in a clear way as a double category (e.g. \cite{johnson20212}), with components:
\begin{itemize}
\item \emph{objects}: event structures.
\item \emph{vertical morphisms $A \to B$}: maps of event structures.
\item \emph{horizontal morphisms $A \profto B$}:  strategies from $A$ to $B$.
\item \emph{2-cells}: maps of strategies. 
\end{itemize}
We define each component in turn.  

\paragraph{Event structures.}
An event structure is a partial order $(A, \leq)$ in which each $a \in A$ has a finite number of predecessors, equipped with a polarity function $A \to \{ {\color{red}-} , {\color{blue}+}\}$, and an irreflexive and symmetric conflict relation $\conflict$ which is hereditary: if $a \leq a'$ and $a \conflict b$ then $a' \conflict b$. The operational intuition is that elements of $A$ are the possible moves in a game, with the polarity function assigning each move to one of the two players.
Moves can only be played when their predecessors in $\leq$ have been played, and moves related by $\conflict$ cannot occur in the same play. 

Therefore the possible states of the game $A$ are the finite subsets $x \subseteq A$ which are down-closed and contain no two moves in conflict; these are called \textbf{configurations}. The set of configurations is written $\Conf{A}$.
For example, the game in \eqref{eq:trivialgame} is an event structure where both $\leq$ and ${\conflict}$ are trivial. 
(Note: we simply call \emph{event structure} what is typically called an \emph{event structure with polarity}.)

\paragraph{Maps of event structures.} A map of event structures $(A, \leq, \conflict) \longrightarrow (B, \leq, \conflict)$ is a function $f: A \to B$ such that for every $x \in \Conf{A}$, $f x \in \Conf{B}$ and moreover $f$ is injective when restricted to $x$. The idea is that any execution of $A$ can be faithfully simulated in $B$. Maps also have to preserve polarity.

\paragraph{Strategies.} A strategy from $A$ to $B$ consists of an event structure $\sigma$ together with a map of event structures
\[
\begin{tikzcd}
\sigma \arrow{r}{\display{\sigma}} &
 A^\perp \parallel B
\end{tikzcd}
\] where the operation $-^\perp$ flips the polarity of every event, and $\parallel$ is a monoidal product representing the parallel composition of event structures. The idea is that every execution allowed by the strategy $\sigma$ must correspond to a play of the game. Strategies must satisfy a technical condition which we omit for now and explain below (Definition~\ref{def:strategy}). 

An important special case is when $A$ is the empty game. Then we say that $\sigma$ is a strategy on the game $B$. For example, all the strategies described in \S\ref{sec:intro:intro} are strategies on the game described in \eqref{eq:trivialgame}, where the arrows $\imc$ correspond to $\leq$. 

\paragraph{Maps of strategies.} 
Let $\sigma$ and $\tau$ be strategies from $A$ to $B$ and from $C$ to $D$, respectively. Suppose that $\alpha: \sigma \to \tau$ makes the following diagram commute, for maps of event structures $f$ and $g$:
\[ 
\begin{tikzcd}
\sigma \arrow{r}{\alpha} \arrow[swap]{d}{\display{\sigma}} &  \tau \arrow{d}{\display{\tau}} \\ 
A^\perp \parallel B  \arrow[swap]{r}{f^\perp \parallel g} & C^\perp \parallel D
\end{tikzcd}
\]  
Then we say $\alpha$ is a map of strategies from $\sigma$ to $\tau$, with boundary $f$ and $g$. These are the 2-cells of the double category.  This double category is weak (or \emph{pseudo}), because the composition of strategies is not strictly associative or unital, only up to invertible 2-cells. 

\begin{remark}
 The specific details of double categories are not important for this paper. Most of this work is only concerned with strategies over a fixed game $A$, their internal symmetries, and notions of maps between strategies over $A$.  
  We only mention the categorical framework in this introduction because it explains the need for uniformity, as we describe next.  
\end{remark}

\subsection{Duplication of moves and uniformity}
\label{sec:intro:uniformity}



The monoidal structure given by parallel composition $\parallel$ is not cartesian: this is a linear model. This is not surprising because  game semantics is closely tied to linear logic. In many applications, for example to give semantics to a higher-order programming language, we resolve this by constructing a resource modality, typically a (pseudo-)comonad $\oc$. Briefly, $\oc A$ is a duplicated version of $A$ in which every move is available in countably many copies. For example, the game in \eqref{eq:trivialgame} is of the form $\oc A$, where $A$ is a game with a single move for each player.
 
But when defining $\oc$ naively we run into issues of uniformity. 
The difficulty is to properly account for the symmetries that arise in the manipulation of copies, as we explain now.

For an event structure $A$, the object $\oc A$ is an infinitary parallel composition:
\[
\oc A = \bigparallel_{i \in \omega} A. 
\]
From any bijection $\alpha : \omega \to \omega^2$, it is possible to construct a family of ``co-multiplication'' strategies $\delta_A : \oc A \profto \oc\oc A$. 
 Unfortunately, for fixed $\alpha$ the associativity law
\begin{equation}
\label{eq:associativity}
\begin{tikzcd}
\oc A \arrow[prof]{r}{ \delta_A} \arrow[swap, prof]{d}{\delta_{A}} & \oc \oc A \arrow[prof]{d}{\delta_{\oc A}} \\ 
\oc \oc A \arrow[prof, swap]{r}{\oc\delta_{ A}} &  \oc\oc \oc A
\end{tikzcd}
\end{equation}
does not hold. We need a way to indicate that the different copies of $A$ in $\oc A$ (and $\oc \oc A$, \emph{etc.}) are exchangeable. This does not make the diagram commute strictly, but it does up to a notion of permutation of moves. 


Permuting moves gives an equivalence relation on strategies. For soundness, this relation must be stable under composition. This requires uniformity: for example, in the non-uniform strategy \eqref{strat4} in \S\ref{sec:intro:intro}, the result of the game is not the same when the opponent Red plays token 1 or token 2.

On the other hand, consider the uniform strategy \eqref{strat1}. If Red decides to update from token 1 to token 2, then the result is the same, although Blue must also permute tokens, to follow the strategy. This is what is called \textbf{bi-invariance} in the work of Melliès \cite{mellies2003asynchronous}: for every Opponent permutation, there is a \emph{response} Player permutation under which the strategy is invariant.


\subsection{Context and related work}

\paragraph*{Uniformity in concurrent games.}

Castellan, Clairambault and Winskel have shown how to address the uniformity problem in concurrent games by upgrading to \emph{event structures with symmetry}, a generalisation introduced by Winskel \cite{winskel2007event} precisely to handle constructions such as $\oc A$. Adding symmetry to concurrent games is challenging and there are several approaches (\cite{castellan2014symmetry,castellan2015parallel}), but a consensus has been reached around a framework known as \emph{thin concurrent games} \cite{cg2}.  
Applications of thin concurrent games are already far-reaching \cite{clairambault2019full,paquetcontinuous}. 

While successful, this construction is technical. Symmetry is often described as a ``proof-relevant equivalence relation'', which indicates which moves are copies of each other via a so-called isomorphism family, \emph{i.e.} a family of bijections between configurations. This means that the construction of the model must account for these families of bijections, both in games and strategies, and this requires additional conditions relating them. 

 We will see how uniform strategies on thin concurrent games arise as a special case of our more general notion of uniform strategies. In particular, we move away from event structures with symmetry and isomorphism families, although we show how they can be recovered (\S\ref{sec:tcg}). 



\paragraph{Uniformity in game semantics in general.}

The need for uniformity arises in any game model which makes duplication explicit. For instance, the pioneering game semantics of Abramsky, Jagadeesan, and Malacaria \cite{abramsky2000full} enforce uniformity via a partial equivalence relation on strategies. This method is too restrictive for the concurrent strategies we consider. 

Many models are based on \emph{plays with pointers}, following Hyland and Ong \cite{hyland2000full}. The uniformity problem does not arise there, because with pointers one can avoid explicit duplication. One drawback is that the underlying linear structure is harder to see. Melliès explains this phenomenon in \cite{mellies2003asynchronous}, motivating his work on asynchronous games; another explanation is given by Castellan and Clairambault in the preprint \cite{DBLP:journals/corr/abs-2103-15453}, which also connects with thin concurrent games. 

There is a well-known intersection of ideas between asynchronous games and thin concurrent games, but an important difference is in the treatment of uniformity. These two lines of research are brought closer together as a by-product of this paper.

\subsubsection*{Outline of the paper.}

In \S\ref{sec:games} we study symmetry in games in terms of two compatible groups actions. This is based on existing ideas \cite{mellies2003asynchronous,cg2}, but the presentation in terms of distributive laws is new to this paper.
In $\S\ref{sec:strategies}$, we explain how to reindex moves in strategies using so-called weak maps between them. Then, in $\S\ref{sec:uniformity}$ we introduce uniform strategies, after developing some algebraic principles for bi-invariance. In $\S\ref{sec:copycat}$ we focus on the copycat strategy, a cornerstone of game semantics, and show that it is uniform. Finally, in $\S\ref{sec:tcg}$, we make a formal connection between the contributions of this paper and the established theory of thin concurrent games.

\section{Games and permutations}
\label{sec:games}

Our first step is to consider games (\emph{i.e.} event structures) equipped with algebraic structure encoding the symmetries. Our presentation uses the definition of a group as a set $G$ equipped with maps 
\[
1 \xrightarrow{e} G \qquad \qquad G \times G \xrightarrow{m} G \qquad \qquad G \xrightarrow{\mathit{inv}} G
\]
satisfying unit, associativity, and inverse laws. We will also use a basic fact about groups:  
\begin{lemma}
For any group $G$, the functor $G \times (-) : \Set \to \Set$ has a canonical monad structure, induced by the maps $m$ and $e$. An algebra over this monad is a set with a left action of $G$.
\end{lemma}

More concretely, a left action of $G$ on a set $A$ is a map $\mathit{act} : G \times A \to A$ such that the diagrams
\[
\begin{tikzcd}
    {1 \times A} \\
    {G\times A} & A
    \arrow["\mathit{act}"', from=2-1, to=2-2]
    \arrow["{e \times A}"', from=1-1, to=2-1]
    \arrow["\cong", from=1-1, to=2-2]
\end{tikzcd}
\qquad \qquad 
\begin{tikzcd}
    {(G\times G) \times A} && {G \times A} \\
    {G \times (G \times A)} & {G \times A} & A
    \arrow["{m \times A}", from=1-1, to=1-3]
    \arrow["\mathit{act}", from=1-3, to=2-3]
    \arrow["\cong"', from=1-1, to=2-1]
    \arrow["{G \times \mathit{act}}"', from=2-1, to=2-2]
    \arrow["\mathit{act}"', from=2-2, to=2-3]
\end{tikzcd}
\]
commute. 
Every group element $g \in G$ induces an automorphism $\mathit{act}(g, -)$ of the set $A$. Indeed an action is equivalently defined as a group homomorphism $G \to \Aut{A}$, where $\Aut{A}$ is the automorphism group of $A$. This definition makes sense for $A$ an object of any category. The action is called \emph{faithful} if distinct elements induce distinct automorphisms, and in this case the group $G$ can be identified with a subgroup of $\Aut{A}$. 

\subsection{Positive and negative automorphisms}

We will consider group actions on games. If $A$ is a game, an automorphism of event structures $\theta : A \to A$ describes a way to permute moves while preserving the dependency and conflict structure.

Two configurations of $A$ are considered symmetric when the action of group element substitutes one with the other. A game can have different symmetry structures, and so we can interpret the same game in different ways. For example, there are exactly two automorphisms of the game $A$ below: the identity map, and the map swapping the left and right columns.
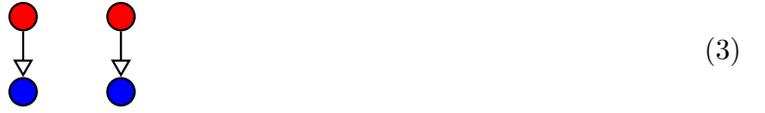
\begin{equation}
\label{eq:gamewithswap}
\raisebox{-2em}{
\begin{tikzpicture}
\node[negative=] (a) at (1, 2) {};
\node[positive=] (b) at (1, 1) {};
\node[negative=] (a') at (2.3, 2) {};
\node[positive=] (b') at (2.3, 1) {};

\draw[control] (a) to (b);
\draw[control] (a') to (b');
\end{tikzpicture}
}
\end{equation}
If the swapping bijection is allowed, then this game is understood as consisting of two exchangeable copies of the same game. Otherwise, $A$ is just a game of the form $B \parallel B$.

In this paper, the symmetry structure on a game $A$ will consist of two groups $\negauto{A}$ and $\posauto{A}$, both acting on $A$ on the left. The idea is that elements of $\negauto{A}$ represent permutations of negative (Red, Opponent) moves, and elements of $\posauto{A}$ are permutations of positive (Blue, Player) moves. Formally the situation is more subtle: since automorphisms of $A$ must preserve the causal structure, in general negative automorphisms have a non-trivial action on positive moves, and vice-versa.

For configurations $x, y$ of an event structure $A$, we write $x \subseteq^+ y$ if $x\subseteq y$ and all moves in $y \setminus x$ are positive; $\subseteq^-$ is defined similarly. 

\begin{definition}
Let $A$ be an event structure with polarity. An automorphism $\alpha \in \Aut{A}$ is \textbf{negative} if it satisfies the following condition: for every $x \in \Conf{A}$, if $\alpha$ fixes $x$, and $x \subseteq^+ y$, then $\alpha$ fixes $y$. Similarly, $\alpha$ is \textbf{positive} if, whenever $\alpha$ fixes $x$, and $x \subseteq^- y$, then $\alpha$ fixes $y$.

An action of a group $G$ on $A$ is \textbf{negative} (resp. \textbf{positive}) if every group element induces a negative (resp. positive) automorphism. 
\end{definition}

\begin{example}






The swapping bijection for the game in \eqref{eq:gamewithswap} is negative, but not positive: it fixes the empty configuration $\emptyset$, but extensions $\emptyset \subseteq^- x$ are not fixed. Note that the Player moves are swapped too, but intuitively this is forced by Opponent. 






\end{example}

Next we define games. The purpose of the definition is to axiomatize the interaction between positive and negative symmetries in the game. 

\subsection{Games equipped with group actions}


\begin{definition}
\label{def:game}
A \textbf{game} is an event structure $A$ equipped with:
\begin{itemize}
 \item a group $\neg$ and a negative group action $\nact : \neg \times A \to A$,
\item a group $\pos$ and a positive group action $\pact : \pos \times A \to A$,
 \item and a distributive law $\lambda : \neg \times \pos \to \pos \times \neg$ between the monads $\neg \times (-)$ and $\pos \times (-)$ on $\Set$,
\end{itemize}
such that the actions are permuted by $\lambda$, as in the diagram below:
\[\begin{tikzcd}
    {(\neg \times \pos) \times A} && {(\pos\times \neg) \times A} \\
    {\neg \times (\pos \times A)} && {\pos \times (\neg \times A)} \\
    {\neg \times A} & A & {\pos \times A}.
    \arrow["{\neg \times \pact}"', from=2-1, to=3-1]
    \arrow["{\lambda \times A}", from=1-1, to=1-3]
    \arrow["\cong", from=1-3, to=2-3]
    \arrow["\pos \times \nact", from=2-3, to=3-3]
    \arrow["\nact"', from=3-1, to=3-2]
    \arrow["\pact", from=3-3, to=3-2]
    \arrow["\cong"', from=1-1, to=2-1]
\end{tikzcd}
\]
\end{definition}

\paragraph*{Distributive laws between groups}
We make some comments about this definition. A distributive law between monads $S$ and $T$ is usually defined as a natural transformation $\lambda_X : ST(X) \to TS(X)$, compatible with the monad structures as specified by four axioms \cite{beck1969distributive}. 

A natural transformation $\neg \times (\pos \times X) \to \pos \times (\neg \times X)$ is determined by its component at $X = 1$, and so it can be presented as a combinator $\lambda : \neg \times \pos \to \pos \times \neg$, as we have done in the definition. The axioms for a distributive law can then be given directly in terms of the group structure in $\neg$ and $\pos$:  
\[\begin{tikzcd}
    {1\times \pos} & {\pos \times 1} & {\neg \times 1} & {1 \times \neg} \\
    {\neg \times \pos} & {\pos \times \neg} & {\neg \times \pos} & {\pos \times \neg}
    \arrow["\cong", from=1-1, to=1-2]
    \arrow["{e \times \pos}"', from=1-1, to=2-1]
    \arrow["{\pos \times e}", from=1-2, to=2-2]
    \arrow["{\lambda }"', from=2-1, to=2-2]
    \arrow["\cong", from=1-3, to=1-4]
    \arrow["{\neg \times e}"', from=1-3, to=2-3]
    \arrow["{\lambda }"', from=2-3, to=2-4]
    \arrow["{e \times \neg}", from=1-4, to=2-4]
\end{tikzcd}\]
\[\begin{tikzcd}[column sep=2em, row sep=1em]
    &  {\neg \times (\pos \times \neg)}  \\
    {\neg \times (\neg  \times \pos)} && {(\neg \times \pos) \times \neg} \\
    {(\neg \times \neg) \times \pos} && {(\pos \times \neg) \times \neg} \\
    && {\pos \times (\neg \times \neg)} \\
    {\neg \times \pos} && {\pos \times \neg}
    \arrow["{\neg \times \lambda }", from=2-1, to=1-2]
    \arrow["\lambda"', from=5-1, to=5-3]
    \arrow["{\lambda \times \neg}", from=2-3, to=3-3]
    \arrow["\cong", from=1-2, to=2-3]
    \arrow["{\pos \times m}", from=4-3, to=5-3]
    \arrow["\cong", from=3-3, to=4-3]
    \arrow["{m \times \pos}"', from=3-1, to=5-1]
    \arrow["\cong"', from=2-1, to=3-1]
\end{tikzcd}\quad
\begin{tikzcd}[column sep=0em, , row sep=1em]
    & {(\pos \times \neg) \times \pos} \\
    {(\neg \times \pos) \times \pos} && {\pos \times (\neg \times \pos)} \\
    {\neg \times (\pos \times \pos)} && {\pos \times (\pos \times \neg)} \\
    && {(\pos \times \pos) \times \neg} \\
    {\neg \times \pos} && {\pos \times \neg}
    \arrow["{\lambda \times \pos}", from=2-1, to=1-2]
    \arrow["\lambda"', from=5-1, to=5-3]
    \arrow["{\pos \times \lambda}", from=2-3, to=3-3]
    \arrow["\cong", from=1-2, to=2-3]
    \arrow["{m \times \neg}", from=4-3, to=5-3]
    \arrow["\cong", from=3-3, to=4-3]
    \arrow["{\neg \times m}"', from=3-1, to=5-1]
    \arrow["\cong"', from=2-1, to=3-1]
\end{tikzcd}
\]
A combinator $\lambda$ of this kind is also known as a \emph{Zappa-Szép product} of the groups $\neg$ and $\pos$ \cite{szep1950structure}.

\paragraph{Permuting subgroups and factorization.}

When $\neg$ and $\pos$ are subgroups of the same group, for example if the actions on $A$ are faithful and $\neg, \pos$ are seen as subgroups of $\Aut{A}$, then the combinator $\lambda$ is uniquely determined provided $\neg$ and $\pos$ are \emph{permuting subgroups}:
\begin{equation}
\label{eq:permuting}
\text{for every $\alpha \in \neg$ and $\beta \in \pos$, there exist $\alpha' \in \neg$ and $\beta' \in \pos$ such that $\alpha\beta = \beta'\alpha'$.}
\end{equation}
In this case, $\alpha'$ and $\beta'$ are unique since $\neg$ and $\pos$ have trivial intersection, and so we must have $\lambda(\alpha, \beta) = (\beta', \alpha').$ All games of interest in semantics seem to have this property, maybe because in practice many group actions are faithful. 

The property \eqref{eq:permuting} arises in thin concurrent games as a factorization property \cite{cg2}. There is a tight relationship between distributive laws, Zappa-Szép products, and strict factorization systems, which explains this \cite{rosebrugh2002distributive}. It is useful in this paper to have an explicit combinator $\lambda$, as part of the structure of a game. 

Any event structure can be seen as a  game with trivial symmetry. In this case none of the moves in $A$ are exchangeable. More interesting games arise from the constructions we describe next. 

\paragraph*{Dual games.}

Any game $A$ has a dual game $A^\perp$. The underlying event structure is the same, with polarity reversed. The symmetry structure on $A^\perp$ is defined by noticing that for any game the combinator $\lambda : \neg \times \pos \to \pos \times \neg$ gives rise to a combinator $\pos \times \neg \longrightarrow \neg \times \pos$ in a canonical way. We will call $\mathit{isw} : \pos \times \neg \to \neg \times \pos$ the invert-and-swap mapping $(\alpha, \beta) \mapsto (\mathit{inv}(\beta), \mathit{inv}(\alpha))$.  
\begin{lemma}
For a game $A$, the event structure $A^\perp$ equipped with $\negauto{A^\perp} = \posauto{A}$, $\posauto{A^\perp} = \negauto{A}$, and the combinator
\[
\lambda_{A^\perp} = 
\begin{tikzcd}
\pos_A \times \neg_A \arrow{r}{\mathit{isw}} & \neg_A \times \pos_A \arrow{r}{\lambda} & \pos_A \times \neg_A \arrow{r}{\mathit{isw}} & \neg_A \times \pos_A, 
\end{tikzcd}
\] 
is a game known as the \textbf{dual} of $A$. 
\end{lemma}

\paragraph*{Parallel composition: a tensor product of games.}
Two games $A$ and $B$ can be combined using the parallel composition of event structures:

\begin{definition}
For event structures $A$ and $B$, the \textbf{parallel composition} $A \parallel B$ is the event structure with events $A \uplus B$, and with $\leq,$ $\conflict$, and $\pol$ inherited from $A$ and $B$.
\end{definition}

Since there is no conflict between events of $A$ and $B$, every configuration of $A \parallel B$ is of the form $x_A \parallel x_B$, and so there is a canonical isomorphism $\Conf{A\parallel B} \cong \Conf{A} \times \Conf{B}.$  
We use this to define the symmetry structure in $A \parallel B$. 

\begin{lemma}
Define the \textbf{parallel composition} of games $A$ and $B$ as the event structure $A \parallel B$, equipped with groups $\neg_{A \parallel B} = \neg_A \times \neg_B$ and $\pos_{A \parallel B} = \pos_A \times \pos_B$ having the product action on $\Conf{A \parallel B} \cong \Conf{A} \times \Conf{B}$. The combinator $\lambda_{A \parallel B}$ is given by  
\[
\neg_{A \parallel B} \times \pos_{A \parallel B}
\xrightarrow{\cong}
(\neg_A \times \pos_A) \times (\neg_B \times \pos_B)
\xrightarrow{\lambda_A \times \lambda_B} 
(\pos_A \times \neg_A) \times (\pos_B \times \neg_B)
\xrightarrow{\cong}
\pos_{A \parallel B} \times \neg_{A \parallel B}.
\] 
This satisfies the axioms for a game. 
\end{lemma}

\paragraph*{A resource modality.}

In game semantics, types are interpreted as games whose initial moves all have the same polarity, and depending on this polarity we call games of this kind either \emph{negative} or \emph{positive}. The idea is that, for a negative game $A$, the copies of $A$ in $\oc A$ can be permuted only by elements of $\neg_{\oc A}$. 

\begin{lemma}
For a negative game $A$, the event structure $\oc A = \bigparallel_{i \in \omega} A$ is equipped with groups 
\begin{align*}
\neg_{\oc A} &= \{ (\pi, (\alpha_i)_{i\in \omega}) \mid \pi \text{ is a permutation of $\omega$, and } \alpha_i \in \neg_A \text{ for all }i \in \omega \} 
\\ 
\posauto{\oc A} &= \{ (\alpha_i)_{i \in \omega} \mid \alpha_i \in \pos_A \text{ for all }i \in \omega \} 
\end{align*}  
under componentwise multiplication, with action on $\oc A$ defined by
\begin{align*}
\nact((\pi, (\alpha_i)_{i\in \omega}), (i, a)) &= (\pi(i), \alpha_i(a)) \\
\pact((\alpha_i)_{i\in \omega}, (i, a)) &= (i, \alpha_i(a))
\end{align*}
and distributive law 
\begin{align*}
\lambda_{\oc A} : \neg_{\oc A} \times \pos_{\oc A} &\longrightarrow \pos_{\oc A} \times \neg_{\oc A} \\ 
((\pi, (\alpha_i)_{i \in \omega}), (\beta_i)_{i \in \omega}) &\longmapsto \left(\left(\beta'_{\pi^{-1}(i)}\right)_{i \in \omega}, (\pi, (\alpha'_i)_{i \in \omega})\right) 
\end{align*}
where for every $i \in \omega$, $(\beta'_i, \alpha'_i) = \lambda_A(\alpha_i, \beta_i).$
This satisfies the axioms for a game. 
\end{lemma}

Other important constructions are involved in applications of game semantics, including a cartesian product $A \with B$ and a linear function space $A \multimap B$ (e.g. \cite{cg2}). 
We omit them here, because the focus is on algebraic foundations rather than applications, and these constructions do not interfere with the group actions. 


\subsection*{Global and local reindexing maps}

We discuss a point of notation. The group actions associated to a game $A$ induce automorphisms of the game. For $\alpha \in \neg$, we will write $\alpha : A \to A$ for the induced automorphism. This is an abuse of notation but  should cause no confusion. Such a map $\alpha$ is a \emph{global} reindexing of the moves of $A$. 

We often also need to discuss the \emph{local} reindexing of a specific configuration, and so for $x \in \Conf{A}$ we will write $\alpha : x \to y$ to mean the restriction of $\alpha$ to $x$, where $y = \alpha x$ is the image of $x$. We will do this for any map of event structures. Note that the local  restriction is always a bijection, because maps of event structures must be locally injective. 



\section{Strategies and permutations}
\label{sec:strategies}

We consider strategies on a fixed game $A$. The definition is due to Rideau and Winskel \cite{rideau2011concurrent}:
\begin{definition}
\label{def:strategy}
A \textbf{strategy} on a game $A$ consists in an event structure $\sigma$, together with a projection map $p_\sigma : \sigma \to A$ such that, for every $x \in \Conf{\sigma}$, 
\begin{itemize}
\item if $p_\sigma x \subseteq^- z$, there is a unique $y \in \Conf{\sigma}$ such that $x \subseteq y$ and $p_\sigma y = z$, and 
\item if $z \subseteq^+ p_\sigma x$, there is a (necessarily unique) $y \in \Conf{\sigma}$ such that $y \subseteq x$ and $p_\sigma y = z$.
\end{itemize}
We write $\sigma : A$ to mean that $\sigma$ is a strategy on $A$. 
\end{definition}
These axioms do not play a major role in this paper: informally they say that a strategy (for Player) should not restrict the behaviour of Opponent or the order in which Opponent processes the Player moves.  Next we define a basic notion of maps between strategies \cite{rideau2011concurrent}:  
\begin{definition}
A \textbf{strict map of strategies} from $\sigma : A$ to $\tau : A$ is a map of event structures $f : \sigma \longrightarrow \tau$ which commutes with the projection maps:
$$
\begin{tikzcd}[column sep=1em]
\sigma \arrow[swap]{dr}{\display{\sigma}} \arrow{rr}{f} & & \tau \arrow{dl}{\display{\tau}} \\ 
& A &  
\end{tikzcd}
$$
\end{definition}
We will now give a more general notion of map. With a strict map as above, we have that $\display\sigma x = \display{\tau} f x$ for every configuration $x \in \Conf{\sigma}$. Since our games now have symmetry, we can relax this equality and allow for a local reindexing of Player moves via a positive bijection $f[x]$. 

\begin{definition}
A \textbf{weak map of strategies} from $\sigma : A$ to $\tau : A$ consists of a map $f : \sigma \longrightarrow \tau$ together with, for every $x \in \Conf{\sigma}$, a positive automorphism $f[x] \in \pos$ such that the following diagram commutes:
\[
\begin{tikzcd}
x \arrow{r}{f} \arrow[swap]{d}{\display\sigma} & fx \arrow{d}{\display\tau} \\ 
\display\sigma x & \arrow{l}{f[x]} \display{\tau} f x.
\end{tikzcd}
\]
\end{definition}
This gives a generalized notion of maps, with strict maps as a special case. Importantly, two strategies are now isomorphic when the underlying event structures are isomorphic and the projections  agree up to a Player permutation. (This would allow us to define an associativity 2-cell filling the diagram \eqref{eq:associativity}.)

\begin{example}
\label{ex:globalweak}
The simplest examples of non-strict weak maps of strategies are those $f : \sigma \to \tau$ for which $f[x] = f[y]$ for every $x, y \in \Conf{\sigma}$. This means that the Player reindexing is \emph{global}, and  we then have $p_\sigma = \beta \circ p_\tau \circ f$ for some $\beta \in \pos$. (Conversely, if the latter is true, then $f$ is a weak map of strategies with every $f[x] = \beta$.)
\end{example}

\newcommand{\Strat}{\mathrm{Strat}}
We write $\Strat(A)$ for the category of strategies on $A$ and weak maps between them, with identities and composition defined in terms of those in $\pos$. The next lemma is an easy observation that will be important for the next section. 
\begin{lemma}
\label{lem:kleisli}
There is a functor $\Strat(A) \longrightarrow \Set_{\pos \times (-)}$, where $\Set_{\pos \times (-)}$ is the Kleisli category for the monad $\pos \times (-)$, that sends a strategy $\sigma$ to the set $\Conf{\sigma}$ and a weak map $f : \sigma \to \tau$ to the Kleisli function
\begin{align*}
\Conf{\sigma} &\longrightarrow \pos \times \Conf{\tau} \\ 
x &\longmapsto (f[x], fx).
\end{align*}
\end{lemma}
This functor is faithful, although not injective on objects.

\section{Uniform strategies}
\label{sec:uniformity}

We introduce our notion of uniformity. The approach is guided by the suggestion (\cite{mellies2003asynchronous}) that uniformity can be understood as a group-theoretic property. 
Informally, a uniform strategy should be invariant under any permutation of Opponent moves, \emph{up to} a permutation of Player moves. We show how to express this formally.

\subsection{Permutations of Opponent moves}

Let $\sigma : A$ be a strategy. For every $\alpha \in \neg$, we can build a new strategy, that has the same underlying event structure $\sigma$ but with projection map 
$$ \sigma \overset{\display{\sigma}}{\longrightarrow} A \overset{\alpha}{\longrightarrow} A.$$
This is a version of $\sigma$ in which moves have been permuted according to $\alpha$. 
This operation forms a left action of $\neg$ on the set of strategies on $A$. 
This new strategy (the result of $\alpha$ acting on $\sigma$) is denoted $\alpha \act \sigma$. 

The goal of uniformity is to ensure that this action leaves the strategy unchanged, except for a relabelling of Player moves as permitted by $\pos$.
One challenge is that the Player response to some Opponent permutation $\alpha \in \neg$ may not be consistent across the strategy, and so we need to reason locally. 

To enforce uniformity, we will ask for a weak map of strategies $\phi_\alpha : \alpha \cdot \sigma \to \sigma$, for every $\alpha \in \neg$. Each $\phi_\alpha$ determines a map 
\begin{equation}
\label{eq:actionresponse}
\Conf{\sigma} \longrightarrow \pos \times \Conf{\sigma} 
\end{equation}
which records the image of every configuration and a suitable local response to $\alpha$. (As a consequence, the action of $\neg$ on strategies extends to a functorial operation on weak maps. This is not true for weak maps between non-uniform strategies.) 

\subsection{Algebras for bi-invariance}

We discuss an algebraic formalization of bi-invariance with respect to two groups and a distributive law between them. For this section, assume $\neg, \pos$ and $\lambda$ are fixed and arbitrary. 

Recall (\S\ref{sec:games}) that sets with a left action of the group $\neg$ are precisely algebras for the monad $\neg \times (-)$ on $\Set$.
To account for the Player response in $\eqref{eq:actionresponse}$, the key idea is to move to the Kleisli category $\Set_{\pos \times (-)}$.

We recall the following basic fact about distributive laws (\cite{beck1969distributive}): 
\begin{lemma}
If $\pos$ and $\neg$ are groups with a distributive law $\lambda : \neg \times \pos \to \pos \times \neg$, the monad $\neg \times (-)$ on $\Set$ lifts to a monad on the Kleisli category $\Set_{\pos \times (-)}.$  
\end{lemma}
An algebra over the lifted monad consists of a set $X$ together with a map $\neg \times X \to \pos \times X$. This is what we take as a basic model for the bi-invariance of $X$ with respect to $\neg$ and $\pos$. 

The functor part of this monad takes a set $X$ to $\neg \times X$, and a Kleisli map $X \xrightarrow{f} \pos \times Y$ to the map $\neg \times X \to \pos \times (\neg \times Y)$ obtained by post-composing $\neg \times f$ with $\lambda \times Y$, modulo associativity isomorphisms. Rather than explaining the full monad structure, we give the axioms for its algebras:
\begin{lemma}
A pair $(X,\ \neg \times X \xrightarrow{h} \pos \times X)$ is an algebra over the monad $\neg \times (-)$ on $\Set_{\pos \times (-)}$ whenever the following equations hold in $\Set$, writing $(\mu^\pos, \eta^\pos)$ and $(\mu^\neg, \eta^\neg)$ for the canonical monad structures: 
\[
\begin{tikzcd}[row sep=1.2em]
\neg\times (\neg \times X) \arrow[swap]{dd}{\mu^\neg_X} \arrow{r}{\neg \times h} & \neg \times (\pos \times X) \arrow{r}{\lambda_X} & \pos \times (\neg \times X) \arrow{d}{\pos \times h} \\ 
 & & \pos \times (\pos \times X) \arrow{d}{\mu^\pos_X} \\ 
\neg \times X \arrow[swap]{rr}{h} & & \pos \times X 
\end{tikzcd}
\qquad\qquad  
\begin{tikzcd}
X\arrow[swap]{d}{\eta^\neg_X}\arrow{drr}{\eta^\pos_X} & & \\[2em]
\neg \times X \arrow[swap]{rr}{h} & & \pos \times X
\end{tikzcd}
\]
\end{lemma}  

Next we use this algebraic structure to define uniform strategies. 

\subsection{Uniform strategies}

\begin{definition}
A \textbf{uniform strategy} on a game $A$ is a strategy $\sigma$ equipped with a weak map 
$
\phi_\alpha : \alpha \cdot \sigma \to \sigma 
$ 
for every $\alpha \in \neg$, such that the induced map 
\[ 
\phi : \neg \times \Conf{\sigma} \longrightarrow \pos \times \Conf{\sigma}
\]
defined by $\phi(\alpha, x) = (\phi_\alpha[x], \phi_\alpha x)$ makes $(\Conf{\sigma}, \phi)$ an algebra over the monad $\neg \times (-)$ on $\Set_{\pos \times (-)}$.

\end{definition} 

Since $\phi$ determines the family of weak maps $(\phi_\alpha)_{\alpha \in \neg}$, we denote a uniform strategy by a pair $(\sigma, \phi).$  It is helpful to unfold the definition. For any $\alpha \in \neg$ and $x \in \Conf{\sigma}$, we have a diagram 
\[
\begin{tikzcd}
    x && y \\
    {p_\sigma x } & {\alpha p_\sigma x} & {p_\sigma y}
    \arrow["\alpha"', from=2-1, to=2-2]
    \arrow["{\phi_\alpha}", from=1-1, to=1-3]
    \arrow[from=1-3, to=2-3]
    \arrow["{\phi_\alpha[x]}", from=2-3, to=2-2]
    \arrow[from=1-1, to=2-1]
\end{tikzcd}
\]
which must commute because $\phi_\alpha$ is a weak map. To unfold the composition axiom, consider $\alpha, \alpha' \in \neg$. We apply the distributive law, to get $(\gamma, \beta) = \lambda(\alpha', \phi_\alpha[x])$, and the overall situation is depicted as
\[
\begin{tikzcd}
    x && y && z \\
    {p x } & {\alpha p x} & {p y} & {\beta p y} & {p z} \\
    && {\alpha'\alpha px }
    \arrow["\alpha", from=2-1, to=2-2]
    \arrow["{\phi_\alpha}", from=1-1, to=1-3]
    \arrow[from=1-3, to=2-3]
    \arrow["{\phi_\alpha[x]}"', from=2-3, to=2-2]
    \arrow[from=1-1, to=2-1]
    \arrow["{\phi_\beta}", from=1-3, to=1-5]
    \arrow["\beta", from=2-3, to=2-4]
    \arrow[""{name=0, anchor=center, inner sep=0}, "{\alpha'}"', from=2-2, to=3-3]
    \arrow[""{name=1, anchor=center, inner sep=0}, "\gamma", from=2-4, to=3-3]
    \arrow["{\phi_{\alpha'\alpha}}", curve={height=-24pt}, from=1-1, to=1-5]
    \arrow[from=1-5, to=2-5]
    \arrow["{\phi_\beta[y]}"', from=2-5, to=2-4]
    \arrow["{\phi_{\alpha'\alpha}[x]}", curve={height=-12pt}, from=2-5, to=3-3]
    \arrow["{\alpha'\alpha}"', curve={height=12pt}, from=2-1, to=3-3]
    \arrow["\lambda"{description}, shift left=1, color={rgb,255:red,214;green,92;blue,92}, shorten <=7pt, shorten >=7pt, Rightarrow, from=0, to=1]
\end{tikzcd}
\]
where the top component of the diagram must commute, and $\gamma \circ \phi_\beta[y] = \phi_{\alpha' \alpha}[x]$, so that the bottom-right triangle commutes.

We now ensure that weak maps are compatible with the uniformity structure. 

\begin{definition}
A \textbf{weak map of uniform strategies} from $(\sigma, \phi)$ to $(\tau, \psi)$ is a weak map $f : \sigma \to \tau$ such that the induced map $\Conf{\sigma} \to \pos \times \Conf{\tau}$ is an algebra homomorphism $(\Conf{\sigma}, \phi) \to (\Conf{\tau}, \psi)$.
\end{definition}

\newcommand{\UStrat}{\mathrm{UStrat}}

Algebra homomorphisms are closed under composition, and so there is a category $\UStrat(A)$ of uniform strategies over a game $A$, and a functor into the category of algebras. Summarizing, we have
\[
\begin{tikzcd}
\UStrat(A) \arrow[swap]{d}{} \arrow{r}{\Conf{-}} & \arrow{d}{} \Set^{\neg \times (-)}_{\pos \times (-)} \\ 
\Strat(A) \arrow{r}{\Conf{-}} \arrow[swap]{r}{(\emph{cf.} \ \ref{lem:kleisli})} & \Set_{\pos \times (-)}
\end{tikzcd}
\]
where the vertical arrows are forgetful functors. 

\section{The copycat strategy: definition and applications}
\label{sec:copycat} 

The copycat strategy is the identity morphism on a game $A$. It is a strategy $A \to A$, so a strategy on the game $A^\perp \parallel A$. The idea is to play a positive move on one side as soon as the corresponding negative move has been played on the other \cite{rideau2011concurrent}:
\begin{definition}
The event structure $\cc_A$ has the same events as $A^\perp \parallel A$, and the same polarity and conflict relation, but $\leq_{\cc_A}$ is the transitive closure of the relation 
$
{\leq_{A^\perp \parallel A}} \cup  \{ ((0, a), (1, a)) \mid \pol_A(a) = + \} \cup \{ ((1, a), (0, a)) \mid \pol_A(a) = - \}.
$

The \textbf{copycat strategy} on $A$ is the event structure $\cc_A$ equipped with the map $p_{\cc_A} : \cc_A \to A^\perp \parallel A$ with identity action on events. 
\end{definition}

The $\cc$ construction extends to a functorial operation: for every map of event structures $f : A \to B$, the map $f^\perp \parallel f : A^\perp \parallel A \to B^\perp \parallel B$
can also be seen as a map $\cc_f : \cc_A \to \cc_B$.

\subsection{The uniform copycat strategy} 

We equip the copycat strategy on $A$ with a map 
\[ 
\phi : \neg_{A^\perp \parallel A} \times \Conf{\cc_A} \longrightarrow \pos_{A^\perp\parallel A} \times \Conf{\cc_A}
\]
as required for uniformity. The construction of $\phi$ is straightforward using the axioms for a game. Let $(\beta, \alpha) \in \neg_{A^\perp \parallel A} = \pos_A \times \neg_A$. The key observation is that $\beta$ and $\alpha$ can be made to agree using the distributive law: if $(\gamma^{-1}, \delta) = \lambda(\alpha, \beta^{-1})$, then (regarding group elements as automorphisms) the diagram 
\[
\begin{tikzcd}[row sep=-0.3em]
& A \arrow{dr}{\delta} & \\ 
A\arrow{ur}{\beta} \arrow[swap]{dr}{\alpha} & & A \\ 
& A \arrow[swap]{ur}{\gamma}& 
\end{tikzcd}
\]
commutes. We can set $\varepsilon := \delta \circ \beta = \gamma \circ \alpha$, so that the diagram below commutes:
\[ 
\begin{tikzcd}
\cc_A \arrow{rr}{\cc_\varepsilon} \arrow[swap]{d}{p} & & \cc_A  \arrow{d}{p} \\ 
A^\perp \parallel A \arrow[swap]{r}{\beta \parallel \alpha} & A^\perp \parallel A \arrow[swap]{r}{\delta \parallel \gamma} & A^\perp \parallel A 
\end{tikzcd}
\]
By the discussion in Example~\ref{ex:globalweak},  $\cc_\varepsilon$ is a weak map $(\beta, \alpha) \cdot \cc_A \to \cc_A$. We set $\phi_{(\beta, \alpha)} := \cc_\varepsilon.$

\begin{lemma}
The pair $(\cc_A, \phi)$, as defined in the preceding discussion, defines a uniform strategy.  
\end{lemma}

 
\subsection{Lifting maps of event structures to strategies}

Copycat has other applications beyond providing identity morphisms. For certain well-behaved maps of event structures $f : A \to B$, we can use copycat to lift $f$ to a \emph{strategy} $\lift{f} : A \profto B$ (\cite{winskel2017games,paquet2020probabilistic,winskel2022}). 
This is often used to lift basic structural morphisms such as the symmetry of parallel composition $A \parallel B \to B \parallel A$, or a co-multiplication $\oc A \to \oc \oc A$. 

\begin{definition}
\label{def:lifting}
Let $f : A \to B$ be a map of event structures which satisfies the conditions for $A$ to be a strategy on $B$, with projection map $f$. Then the strategy $\lift{f} :  A \profto B$ 
is defined as the event structure $\cc_A$ with projection map 
\[ 
\cc_A \xrightarrow{p_{\cc_A}} A^\perp \parallel A \xrightarrow{A^\perp \parallel f} A^\perp \parallel B.
\] 
\end{definition}
The purpose of this section is to identify sufficient conditions on $f$ under which $\lift{f}$ can be  made uniform. The basic requirement is that $f$ should reflect the negative group action and preserve the positive group action, in a functorial and coherent way. 

\begin{lemma}
\label{lem:liftingsymm}
Let $f : A \to B$ be a map of event structures as in Definition~\ref{def:lifting}, such that there are group homomorphisms $L : \neg_B \to \neg_A$ and $M : \pos_A \to \pos_B$ such that for every $\alpha \in \neg_B$ and $\beta \in \pos_A$, 
\[ 
\begin{tikzcd}
A \arrow[swap]{d}{f} \arrow{r}{L(\alpha)} & A \arrow{d}{f} \\ 
B\arrow[swap]{r}{\alpha} & B 
\end{tikzcd}
\qquad \qquad
\begin{tikzcd}
A \arrow[swap]{d}{f} \arrow{r}{\beta} & A \arrow{d}{f} \\ 
B\arrow[swap]{r}{M(\beta)} & B 
\end{tikzcd}
\]
and additionally the following coherence property holds:
\[\begin{tikzcd}[row sep=0em, column sep=3em]
    & {\neg_B \times \pos_B} & {\pos_B \times \neg_B} \\
    {\neg_B \times \pos_A} &&& {\pos_B \times \neg_A} \\
    & {\neg_A \times \pos_A} & {\pos_A \times \neg_A}
    \arrow["{\neg_B \times M}"{pos=0.8}, from=2-1, to=1-2]
    \arrow["{\lambda_B}", from=1-2, to=1-3]
    \arrow["{\pos_B \times L}"{pos=0.2}, from=1-3, to=2-4]
    \arrow["{L \times \pos_A}"'{pos=0.8}, from=2-1, to=3-2]
    \arrow["{\lambda_A}"', from=3-2, to=3-3]
    \arrow["{M \times \neg_A}"'{pos=0.2}, from=3-3, to=2-4]
\end{tikzcd}\]
Then, the strategy $\lift{f}$ can be made uniform with $\phi_{\lift{f}}$ defined as the composite
\[
\neg_{A^\perp \parallel B} \times \Conf{\cc_A} 
\xrightarrow{(\neg_{A^\perp} \times L) \times \Conf{\cc_A}} 
\neg_{A^\perp \parallel A} \times \Conf{\cc_A} 
\xrightarrow{\phi_{\cc_A}} 
\pos_{A^\perp \parallel A} \times \Conf{\cc_A} 
\xrightarrow{(\pos_{A^\perp} \times M) \times \Conf{\cc_A}}
\pos_{A^\perp \parallel B} \times \Conf{\cc_A}.
\]
\end{lemma}



\begin{remark}
The lifting construction has a \emph{co-lifting} counterpart: in some cases, for a map $f : A \to B$, the composite 
\[ 
\cc_A \xrightarrow{p_{\cc_A}} A^\perp \parallel A \xrightarrow{f^\perp \parallel A} B^\perp \parallel A.
\]
gives a strategy from $B$ to $A$, which can be made uniform providing $f$ is compatible with symmetry in a way that is essentially dual to Lemma~\ref{lem:liftingsymm}.
\end{remark}

A key application of these results is in the definition of a comonad structure for $\oc$. For example the counit $\oc A \profto A$ is obtained by co-lifting the injection $A \to \oc A$. Liftings and co-liftings of this kind have a universal property \cite{shulman2008framed} (see also \cite{paquet2020probabilistic}), which provides canonical constructions for 2-cells between lifted maps, as required for \eqref{eq:associativity}. We do not develop this further, as this paper focuses on the study of uniformity. 

\section{Thin concurrent games}
\label{sec:tcg}

We explore the connection between the algebraic development in this paper and the established theory of thin concurrent games. We recall the main definitions of games, strategies, and maps of strategies from \cite{cg2,castellan2015parallel}, and explain how they arise within our setting.  

\subsection{Symmetry in event structures and games}

First we focus on the games. Permutation of moves in thin concurrent games is described using isomorphism families \cite{winskel2007event}:
\begin{definition}
Let $E$ be an event structure. An \textbf{isomorphism family} on $E$ is a set $\sym{E}$ of polarity-preserving bijections between configurations of $E$, containing identities and closed under composition and taking  inverses, such that for every $(\theta : x \cong y)\in \sym{E}$:
\begin{description}
    \item[\ (restriction)] if $x' \subseteq x$, then the restriction of $\theta$ to $x'$ is in $\sym{E}$; and 
    \item[\ (extension)] if $x \subseteq x'$, then there exists some $\theta' : x' \cong y'$ in $\sym{E}$ such that $\theta$ is the restriction of $\theta'$ to $x$.
\end{description}
The pair $(E, \sym{E})$ is then called an \textbf{event structure with symmetry}. Maps of event structures with symmetry are also required to preserve symmetry. 
\end{definition}

Castellan, Clairambault and Winskel define a notion of game with symmetry \cite{cg2}:
\begin{definition}
A \textbf{thin concurrent game} is an event structure $A$ equipped with three isomorphism families $\sym{A}, \possym{A},$ and $\negsym{A}$ such that $\possym{A}, \negsym{A} \subseteq \sym{A}$, and:
\begin{itemize}
\item if $\theta \in \possym{A} \cap \negsym{A}$, then $\theta = \id_x$ for some $x \in \Conf{A}$;
\item if $\theta \in \possym{A}$ and $\theta \subseteq^+ \theta' \in \sym{A}$, then $\theta' \in \possym{A}$; and 
\item if $\theta \in \negsym{A}$ and $\theta \subseteq^- \theta' \in \sym{A}$, then $\theta' \in \negsym{A}$.
\end{itemize}
\end{definition}

\newcommand{\tcg}[1]{\underline{#1}}
With the next result we begin to explain the connection with the work in this paper. 
\begin{lemma}
\begin{enumerate}
\item 
Let $G$ be a group acting on an event structure $E$. Then the set  
\[ 
\symgp{G}{E} = \{ \theta : x \cong y \mid \theta \text{ is the restriction to $x$ of some $E \xrightarrow{g} E$}, g \in G \}
\]
defines an isomorphism family on $E$. 
\item Let $A$ be a game in the sense of Definition~\ref{def:game}. Then the structure of $A$ defines a thin concurrent game $\tcg{A}$, where $\negsym{\tcg{A}} = \symgp{\neg}{A}$, $\possym{\tcg{A}} = \symgp{\pos}{A}$, and $\sym{\tcg{A}}$ is the closure of $\negsym{\tcg{A}} \cup \possym{\tcg{A}}$ under composition. 
\end{enumerate}
\end{lemma}

We have shown that our games give rise to thin concurrent games. It seems that all useful thin concurrent games arise in this way, although there are pathological examples, which arise because isomorphism families provide bisimulations rather than automorphisms. 

\subsection{Symmetry in strategies}

In thin concurrent games, strategies are also equipped with isomorphism families.

\begin{definition}
Let $A$ be a thin concurrent game. A \textbf{$\sim$-strategy} on $A$ is an event structure with symmetry $\sigma$, together with a map $(\sigma, \sym{\sigma}) \to (A, \sym{A})$ whose underlying map is a strategy, satisfying a further two axioms: 
\begin{description}
\item[(thinness)] If $x \in \Conf{\sigma}$, and $\id_x \subseteq^+ \theta \in \sym{\sigma}$, then $\theta = \id_{x'}$ for some $x' \in \Conf{\sigma}$.
\item[($\sim$-receptivity)] If  $x \subseteq^{-} y, z \in \Conf{\sigma}$, and there exists some $(\theta : py \cong pz) \in \sym{A}$ such that $\id_{p x} \subseteq \theta$, then there exists $\chi \in \sym{\sigma}$ such that $\id_x \subseteq \chi$ and $p\chi = \theta$. 
\end{description}
\end{definition}
As we will see below (Proposition~\ref{prop:localequivalence}), we can explain these axioms by imposing a \emph{locality} requirement on the uniformity.  To achieve this, we connect our uniform strategies to $\sim$-strategies.  In the next lemma, we show that for every uniform strategy $\sigma : A$, it is possible to recover an isomorphism family on $\sigma$. 
\begin{lemma}
Let $(\sigma, \phi)$ be a uniform strategy on a game $A$. Then 
\[
\symgp{\phi}{\sigma} = \{ \theta : x \cong y \mid \theta \text{ is the restriction to $x$ of } \phi_\alpha, \text{ for some } \alpha \in \neg \}
\]
is an isomorphism family on the event structure $\sigma$, and the map $p_\sigma : \sigma \to A$ is a map of event structures with symmetry $(\sigma, \symgp{\phi}{\sigma}) \to (A, \sym{\tcg{A}})$. 
\end{lemma}

Unfortunately, thinness and $\sim$-receptivity do not hold in general, as the next example shows. 
\begin{example}
\label{ex:nonlocal}
Consider the simple game $A$ given by the event structure
\[
\begin{tikzpicture}
\node[negative=0] (0) at (0, 0) {};
\node[negative=1] (1) at (1, 0) {};

\node[positive=0] (0') at (2.5, 0) {};
\node[positive=1] (1') at (3.5, 0) {};
\end{tikzpicture}
\]
where $\neg = \{ \alpha, \id \}$ and $\pos = \{ \beta, \id \}$ are the symmetry groups for the negative and positive events, respectively, with the canonical action, and trivial  distributive law. 

Consider the strategy $\sigma = A$ where $p_\sigma$ is the identity map. We construct a bi-invariance structure $\phi$. By the axioms, $\phi_\id$ must be the identity map on $\sigma$. Define $\phi_\alpha = \beta \circ \alpha$, so that swapping the Opponent moves also triggers a swap of the Player moves. 
This satisfies the axioms for a uniform strategy, since both $\alpha$ and $\beta \circ \alpha$ are their own inverse. 

But the family $\symgp{\phi}{\sigma}$, obtained by restricting $\phi_\alpha$ and $\phi_\id$ to individual configurations, contains a bijection 
\begin{equation}
\label{eq:bijectionnonthin}
\raisebox{-0.9em}{\begin{tikzpicture}\node[positive=0] (0') at (0, 0) {};\end{tikzpicture}}
\quad \quad \cong \quad \quad
\raisebox{-0.9em}{\begin{tikzpicture}\node[positive=1] (0') at (0, 0) {};\end{tikzpicture}}
\end{equation}
which breaks thinness, as a non-identity positive extension of the identity on $\emptyset$. (It is also not $\sim$-receptive, taking $x$ to contain the positive moves, and $y, z$ the two one-event negative extensions of $x$.) 
\end{example}

We give an informal analysis of the problem: by providing only a \emph{local} account of symmetry, the isomorphism family $\symgp{\phi}{\sigma}$ fails to record that the bijection \eqref{eq:bijectionnonthin} comes from $\phi_\alpha$, and forgets that it is only valid when the Opponent moves are swapped too. In this local view, it looks as though the two positive moves are swapped for no reason.  This is resolved by the next result. 
\begin{proposition}
\label{prop:localequivalence}
Let $(\sigma, \phi)$ be a uniform strategy on $A$ such that $\phi$ satisfies the following property:
\begin{description} 
    \item[\quad(uniformity is local.)] For all $x \in \Conf{\sigma}$ and $\alpha \in \neg$, if $\alpha$ fixes $p_\sigma x$, then $\phi_\alpha$ fixes $x$.
\end{description}
Then, $(\sigma, \symgp{\phi}{\sigma})$ is a $\sim$-strategy on $\tcg{A}$, written $\tcg{\sigma}_\phi.$ 
\end{proposition}

In particular, uniformity in Example~\ref{ex:nonlocal} is not local. This way, we have identified the strategies on thin concurrent games with a restricted, well-behaved class of uniform strategies, in which all uniformity information is accessible locally. We note that it is \emph{not} the case that every uniform strategy can be assigned a modified $\phi$ for which uniformity is local. Thus we have a proper generalization. 

\subsection{Weak maps of strategies}

\begin{definition}
Let $A$ be a thin concurrent game and let $\sigma, \tau$ be $\sim$-strategies. A \textbf{weak map of $\sim$-strategies}~\cite{cg2} is a map $f : \sigma \to \tau$ of event structures with symmetry, such that for every $x \in \Conf{\sigma}$ there exists a (necessarily unique) bijection $(\theta : p_\sigma x \cong p_\tau f x) \in \possym{A}$ such that the following diagram commutes:
\begin{equation}
\label{eq:diagramweakmaps}
\begin{tikzcd}
x \arrow[swap]{d}{p_\tau} \arrow{r}{f} & fx \arrow{d}{p_\tau} \\ 
p_\sigma x \arrow[swap]{r}{\theta} & p_\tau f x 
\end{tikzcd}
\end{equation}
\end{definition}

As expected, we obtain this by restricting the weak maps between uniform strategies. 
\begin{lemma}
Let $(\sigma, \phi)$ and $(\tau, \psi)$ be uniform strategies on a game $A$, and let $f : \sigma \to \tau$ be a map of uniform strategies between them. Then: 
\begin{enumerate}
\item $f$ is a map of event structures with symmetry $(\sigma, \symgp{\phi}{\sigma}) \to (\tau, \symgp{\psi}{\tau})$; and
\item for every $x \in \Conf{\sigma}$ there exists $\theta \in \symgp{P}{A}$ such that the diagram in \eqref{eq:diagramweakmaps} commutes.
\end{enumerate}
\end{lemma}

As an immediate corollary, we obtain that if uniformity is local in $(\sigma, \phi)$ and $(\tau, \psi)$, a weak map $f$ of uniform strategies induces a weak map $\tcg{f} : \tcg{\sigma}_\phi \to \tcg{\tau}_\psi$ between $\sim$-strategies on the thin concurrent game $\tcg{A}$.


\section{Concluding remarks}

We have defined a new general notion of uniform strategies on event structures. Our presentation has made explicit the algebraic structures and coherence laws underlying the usual treatment of uniformity in game semantics. The work is specific to concurrent games, but the algebraic perspective could facilitate the transfer of ideas to other kinds of games, e.g. \cite{DBLP:conf/lics/Mellies21}.  

Within the landscape of concurrent games, this development will combine smoothly with weights or annotations on event structures, as required to model other kinds of programming (e.g. \cite{DBLP:journals/entcs/Winskel13,alcolei2019jeux}): it suffices to require that each $\phi_\alpha$ preserves the weights.  

This work should be seen as a preliminary investigation. It remains to determine how far we can take the uniform strategies in this paper, in full generality. In particular, adapting the current methods for composing strategies will require some work, as these rely strongly on the local perspective \cite{cg2}. This is an important challenge, because the main purpose of uniformity (\emph{cf.} \S\ref{sec:intro:uniformity}) is to ensure that permutations of copies interact well with the composition of strategies.

\section*{Acknowledgement}
I am grateful to Pierre Clairambault and Simon Castellan for many discussions on the
     topic of this paper, and to Philip Saville for useful feedback on
     a draft. Thanks also to the anonymous reviewers. I acknowledge
     support from a Royal Society University Research Fellowship.

\bibliographystyle{entics}
\bibliography{mfps}

\end{document}